\documentstyle[12pt,fullpage,epsfig]{article}

\newcommand{\Sec}[1]{Section~\ref{#1}}
\newcommand{\eq}[1]{Eq.\,(\ref{eq#1})}
\newcommand{\fig}[1]{Fig.\,\ref{fig#1}}
\newcommand{\tab}[1]{Table\,\ref{tab#1}}
\newcommand{\cE}{{\cal E}_0}

\title{Computer Model of Quantum Zeno Effect in Spontaneous
Decay with Distant Detector}

\author{Alexander D.\ Panov\\
{\small Skobeltsyn Institute of Nuclear Physics,}\\
{\small Moscow State University, Moscow 119899, Russia}
\thanks{E-mail address: {\tt a.panov@relcom.ru}}
}
\date{}

\begin{document}
%%%%%%%%%%%%%%%%%%%%%%%%%%%%%%%%%%%%%%%%%%%%%%%%%%%%%%%%%%%%%%%%%%%
%\bibliographystyle{prsty}
%%%%%%%%%%%%%%%%%%%%%%%%%%%%%%%%%%%%%%%%%%%%%%%%%%%%%%%%%%%%%%%%%%%
%\psdraft

\maketitle
\sloppy
\begin{abstract}
A numerical model of spontaneous decay continuously monitored by a
distant detector of emitted particles is constructed.  It is shown
that there is no quantum Zeno effect in such quantum measurement if
the interaction between emitted particle and detector is short-range
and the mass of emitted particle is not zero.
\end{abstract}

\begin{center}
PACS numbers: 03.65.Bz\\
Keywords: quantum measurement, quantum Zeno effect, spontaneous
decay, numerical model
\end{center}

\section{Introduction}
\label{INTRO}

The Quantum Zeno Paradox (QZP) is a proposition that evolution of a
quantum system is stopped if the state of system is continuously
measured by a macroscopic device to check whether the system is still
in its initial state \cite{ZEN_SUDARSHAN77A,ZEN_SUDARSHAN77B}. QZP is
a consequence of formal application of von Neumann's projection
postulate to represent a continuous measurement as a sequence of
infinitely frequent instantaneous collapses of system's wave
function. It was shown theoretically \cite{ZEN_COOK88} and
experimentally \cite{ZEN_ITANO90} that sufficiently frequent discrete
active measurements of system's state really inhibit quantum
evolution. This phenomenon was named `Quantum Zeno Effect' (QZE).
But the question about possibility of QZE during {\em true
continuous} observations is not quite clear up to now.

A true continuous measurement of quantum system's state takes place
during observation of spontaneous decay by distant detector of
emitted particles (another examples of continuous measurement of
decay are presented in papers
\cite{ZEN_PASCAZIO97,ZEN_PASCAZIO99A,ZEN_MENSKY99A,%
ZEN_ELATTARI99,ZEN_ELATTARI00}).  Let us consider a metastable exited
atom surrounded for detectors to register an emitted photon (or
electron) when the exited state of atom decays to the ground state.
While the detectors are not discharged, the information that the atom
is in its exited state is being obtained permanently, therefore the
system's state is being measured continuously.  Could the presence of
detectors influence on the decay constant of exited atom?  If so,
this would be the QZE in true continuous passive measurement.

It is impossible to describe this kind of continuous measurement by a
sequence of discrete wave function collapses as was proposed in
seminal works \cite{ZEN_SUDARSHAN77A,ZEN_SUDARSHAN77B}. Such approach
leads to the explicit quantum Zeno paradox, not effect.  Instead, a
dynamical description of such measurements was elaborated in the
number of works \cite{ZEN_KRAUS81,ZEN_SUDBERY84,PANOV96F,PANOV99B}.
In this approach object system (atom), radiation field (or emitted
particle), and device (detector of particle) are considered as
subsystems of one compound quantum system. The results of papers
\cite{ZEN_KRAUS81,ZEN_SUDBERY84} were mainly qualitative. The
explicit expression for decay constant perturbed by given interaction
$W$ of emitted particles with detector was obtained in
\cite{PANOV96F,PANOV99B}. This expression is
\begin{equation}
   \Gamma = 2\pi \int d\omega M(\omega) \Delta(\omega - \cE).
   \label{eq1}
\end{equation}
In \eq{1} $M(\omega)$ is the sum of all transition matrix element
squares related to the same energy of emitted particle $\omega$;
$\cE$ is the expectation value of final energy of emitted particle.
The function $\Delta(\omega - \cE)$ describes the influence of
observation on the decay constant. Without detector, i.~e.\ $W=0$,
the function $\Delta(\omega - \cE)$ transforms to Dirac's
delta-function $\delta(\omega - \cE)$ and \eq{1} transforms to the
Golden Rule \cite{PANOV96F,PANOV99B}.  It was supposed
\cite{PANOV96F} that the meaning of $\Delta(\omega - \cE)$ is an
energy spreading of the final states of decay due to time-energy
uncertainty relation and finite time-life of emitted particle until
scattering on the detector\footnote{This supposition was confirmed in
a case of problem of decay onto an unstable atomic state
\cite{PANOV99B}. This problem is close to problem of observation of
decay by distant detector.}.  Then one can suggest that observation
influences on decay in accordance with the following sequence: The
faster detector, the shorter emitted particle time-life, the wider
$\Delta(\omega - \cE)$, the stronger perturbation of decay constant.

It is clear that strong interactions $W$ is needed to obtain QZE.
Hence, $W$ is essentially nonperturbative in this problem. This
feature determines the main difficulty of calculations of function
$\Delta(\omega - \cE)$ in \eq{1} and, consequently, the perturbed
value of decay constant.  Particularly, in paper \cite{PANOV96F} we
supposed that QZE explains strong inhibition of 76\,eV-nuclear
uranium-235 isomer decay in matrix of silver \cite{U_EK_KOLTSOV89}.
However, we had to restrict the consideration only by a qualitative
analysis of \eq{1} for this case because of difficulties of function
$\Delta(\omega-\cE)$ calculations.

Since it is difficult to study realistic physical systems, it is
reasonable to start with some simplified models to calculate the
function $\Delta$. The aim of the present paper is strict and
complete numerical investigation of \eq{1} for a simple but not
oversimplified model system.  We derive \eq{1}, then introduce
one-dimensional three-particle model of continuous observation of
decay, then describe the numerical computation scheme for this model,
and finally discuss results of calculations.

\section{General considerations}
\label{THEORY}

In this section a derivation of \eq{1} and other formulae to
construct our numerical model are presented.  Derivation of \eq{1} is
simplified in comparison with our previous papers
\cite{PANOV96F,PANOV99B}.

\begin{figure}
   \begin{center}
      \epsfig{file=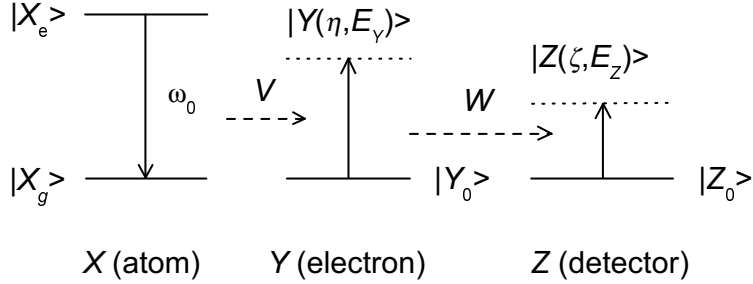,width=10cm}
   \end{center}
   \caption{The compound system $S = X\otimes Y\otimes Z$ is the model
      of continuous observation of exited state $|X_e\rangle$ decay.}
   \label{fig1}
\end{figure}

Let a compound system $S = X\otimes Y\otimes Z$ consists of three
subsystems $X$, $Y$, and $Z$ (\fig{1})\footnote{$S = X\otimes
Y\otimes Z$ means that the Hilbert space of system $S$ is a direct
product of spaces of systems $X,Y,Z$.}.  The system $X$ (``atom'')
decays spontaneously from the initial exited state $|X_e\rangle$ to
the ground state $|X_g\rangle$ emitting a particle $Y$ (``electron'')
due to interaction $V$ between systems $X$ and $Y$. This process is
similar to autoionization decay of excited atomic state, but it is
possible to suppose another nature of systems and interactions.  The
particle $Y$ is initially at the ground state $|Y_0\rangle$
(electron is on the bounded state in atom) and then transits to
continuum $|Y(\eta, E_Y)\rangle$. Here $E_Y$ is the energy of
the state in the continuum and $\eta$ represents all other quantum
numbers.  Particle $Y$ inelastically scatters on third system $Z$
(``distant detector'') due to interaction $W$ between $Y$ and $Z$.
As a result, the system $Z$ transits from the initial ground state
$|Z_0\rangle$ to the continuum $|Z(\zeta,E_Z)\rangle$. This
transition is considered to be a registration of decay. We consider
that the interaction $V$ does not effect on system $Z$, the
interaction $W$ does not effect on system $X$ and the systems $Y$ and
$Z$ don't interact in their ground states.  Therefore, we have
\begin{equation}
   V = V_{XY} \otimes I_Z;\quad
   W = I_X \otimes W_{YZ};\quad
   W_{YZ}|Y_0 Z_0\rangle = 0,
   \label{eq2}
\end{equation}
where $I_X$ and $I_Z$ are unit operators in the Hilbert spaces
of corresponding systems. The Hamiltonian of whole system is
\begin{equation}
   H = H_0 + V + W,
   \label{eq3}
\end{equation}
where
$$
   H_0 = H_X^0 \otimes I_{YZ} +
         H_Y^0 \otimes I_{XZ} +
         H_Z^0 \otimes I_{YX}
   %\label{eq4}
$$
with obvious notations.

The initial state of system $S$ at the initial moment of time
$T=0$ is
$$
   |\Psi_0\rangle = |X_e\rangle \otimes
                    |Y_0\rangle \otimes
                    |Z_0\rangle \equiv
                    |X_e Y_0 Z_0\rangle.
   %\label{eq4Prim}
$$
Let us introduce the first order correction to the eigenenergy of
state $|\Psi_0\rangle$ due to interaction $V$:
$$
   \delta V_0 = \langle\Psi_0|V|\Psi_0\rangle
   %\label{eq5}
$$
and renormalized unperturbed Hamiltonian $H_0$ and renormalized
interaction $V$:
$$
   H_0' = H_0 + \delta V_0|\Psi_0\rangle\langle\Psi_0|;\quad
   V' = V - \delta V_0|\Psi_0\rangle\langle\Psi_0|.
   %\label{eq6}
$$
Then the Hamiltonian \eq{3} may be rewritten as
$$
   H = H_0' + V' + W.
   %\label{eq7}
$$
The initial state $|\Psi_0\rangle$ is an eigenstate of the Hamiltonian
$H_0'$ with the eigenenergy
$$
   \cE' = E^e_X + E^0_Y + E^0_Z + \delta V_0.
   %\label{eq8}
$$

The interaction $V$ is considered to be a small perturbation, but
interaction $W$ is not small. To obtain the decay constant of the
exited state $|X_e\rangle$ it is necessary to solve the Shr\"odinger
equation for the whole system $S$.  It is impossible to construct the
perturbation theory for $W$, but it is possible for $V$. Therefore,
let us introduce the interaction picture as ($\hbar=1$):
\begin{eqnarray}
   |\Psi_I(T)\rangle & = & {\rm e}^{i(H_0'+W)T} |\Psi(T)\rangle,\quad
   |\Psi_I(0)\rangle = |\Psi_0\rangle
   \label{eq9}\\
   V_I'(T) & = & {\rm e}^{i(H_0'+W)T} V' {\rm e}^{-i(H_0'+W)T}.
   \nonumber%\label{eq10}
\end{eqnarray}
Then the Shr\"odinger equation reads as
\begin{equation}
   |\Psi_I(T) = |\Psi_0\rangle -
   i\int_0^T V_I'(t)|\Psi_I(t)\rangle dt.
   \label{eq11}
\end{equation}
The solution of \eq{11} in the second order of perturbation theory
with respect to $V$ is
\begin{equation}
   |\Psi_I(T)\rangle = |\Psi_0\rangle -
   i \int_0^T V_I'(t)|\Psi_0\rangle dt -
   \int_0^T dt_1 \int_0^{t_1} dt_2 V_I'(t_1) V_I'(t_2) |\Psi_0\rangle.
   \label{eq12}
\end{equation}
Let $F(T)$ be no-decay amplitude
$$
   F(T) = {\rm e}^{i\cE'T} \langle \Psi_0 | \Psi(T)\rangle.
   %\label{eq13}
$$
It follows from \eq{9} and \eq{12} that
\begin{equation}
   F(T) = 1 - \int_0^T dt_1 \int_0^{t_1} dt_2
   \langle \Psi_0 | V_I'(t_1) V_I'(t_2) | \Psi_0 \rangle.
   \label{eq14}
\end{equation}
For the initial region of exponential decay curve (time is not very
small, not large) we assume
\begin{equation}
   F(T) = \exp(-\gamma T) \cong 1 - \gamma T,\; \gamma = {\rm const}.
   \label{eq15}
\end{equation}
Then the quantity $\Gamma = 2{\rm Re}\,\gamma$ is the probability of
decay per unit of time (decay constant). Using \eq{14} and
\eq{15}, we obtain
\begin{equation}
   \Gamma = 2 {\rm Re} \int_0^\infty
   \langle \Psi_0 | V'{\rm e}^{-i(H_0'+W)t}V' | \Psi_0 \rangle
   {\rm e}^{i\cE't} dt.
   \label{eq16}
\end{equation}
By $v(\eta, E_Y)$ denote the matrix elements of $V$ which cause the
decay of state $|X_e\rangle$ and emitting of particle $Y$:
\begin{equation}
   v(\eta, E_Y) =
   \langle X_g Y(\eta, E_Y) | V'_{XY} | X_e Y_0 \rangle.
   \label{eq17}
\end{equation}
All other matrix elements don't effect on the decay constant.  Let us
introduce the vector
\begin{equation}
   | \widetilde Y \rangle =
   \int d\eta dE_Y |Y(\eta, E_Y)\rangle v(\eta, E_Y).
   \label{eq18}
\end{equation}
Then, after simple algebraic transformations, \eq{16} may be
rewritten as
\begin{equation}
   \Gamma = 2\pi \int_0^\infty M(E_Y)
   \Delta\left(E_Y - E_Y^{fin}\right) dE_Y,
   \label{eq19}
\end{equation}
where
\begin{eqnarray}
   E_Y^{fin} & = & E_Y^0 + \omega_0 + \delta V_0,\quad
   \omega_0 = E^e_X - E^g_X,
   \nonumber
   %\label{eq20}
   \\
   M(E_Y) & = & \int d\eta \left|v(\eta, E_Y)\right|^2,
   \nonumber
   %\label{eq21}
   \\
   \Delta(E) & = & \frac{1}{\pi}{\rm Re}\int_0^\infty D(t)
   {\rm e}^{-iEt} dt,
   \label{eq22}\\
   D(t) & = &
   \frac
   {
      \langle \widetilde Y Z_0
      \left|
      {\rm e}^{-i(H_{YZ}^0 + W_{YZ})t}
      \right|
      \widetilde Y Z_0 \rangle
   }
   {
      \langle \widetilde Y Z_0
      \left|
      {\rm e}^{-iH_{YZ}^0 t}
      \right|
      \widetilde Y Z_0 \rangle
   },
   \label{eq23}\\
   H_{YZ}^0 & = & H_Y^0 \otimes I_Z + H_Z^0 \otimes I_Y.
   \nonumber
   %\label{eq24}
\end{eqnarray}
\begin{figure}
   \begin{center}
      \epsfig{file=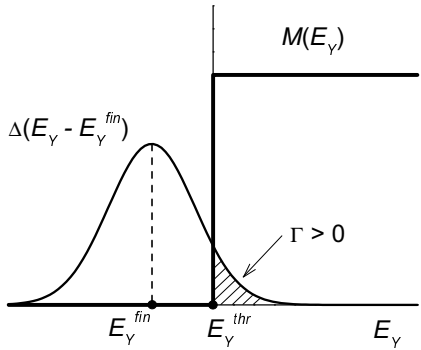,width=8cm}
   \end{center}
   \caption{No-zero probability of decay forbidden by the Energy
   Conservation Law.}
   \label{fig2}
\end{figure}
It is easily shown that $\int \Delta(E) dE = 1$.  We can not obtain
an explicit analytical expression for $\Gamma$ with respect to the
matrix elements of interaction $W$ due to nonperturbative character
of this interaction. However, it is possible to derive an interesting
qualitative conclusion on the shape of function $\Delta(E)$ (which is
essentially used to calculate $\Gamma$) without calculations.  For
simplicity we suppose $M(E_Y)$ to be a step-like function with the
jump at energy $E_Y^{thr}$.  Suppose the energy-spreading function
$\Delta(E)$ be a bell-like and nearly symmetric with the maximum at
$E=0$.  Consider the case $E_Y^{fin} < E_Y^{thr}$ (\fig{2}).   Then
the transition $|X_e\rangle \to |X_g\rangle$ with emitting of
particle $Y$ is strictly forbidden by the Energy Conservation Law.
For example, the binding energy of atomic electron is greater than
the transition energy $\omega_0$ and the electron can not be ionized
during this transition.  But the functions $\Delta(E_Y - E_Y^{fin})$
and $M(E_Y)$ may have no-zero overlap integral \eq{19} as is shown on
\fig{2}.  Hence $\Gamma > 0$ and the transition $|X_e\rangle \to
|X_g\rangle$ is possible.  Thus, we come to a contradiction. This
contradiction means that the suggestion about shape of function
$\Delta(E)$ was wrong.  The contradiction could be eliminated if
whole no-zero part of function $\Delta(E)$ locates at the left-hand
side of point $E=0$. Therefore, we conclude, that our formalisms
predict this special shape for the function $\Delta(E)$.  Another
prediction is the following.  Let $E_Y^{fin} - E_Y^{thr} > 0$ but the
value $E_Y^{fin} - E_Y^{thr}$ is of the order of function $\Delta(E)$
width or less.  Then the transition $|X_e\rangle \to |X_g\rangle$ is
permitted, but a considerable part of function
$\Delta(E_Y-E_Y^{fin})$ is located at the left-hand side of point
$E_Y^{thr}$, so $\Gamma < \Gamma_0$.  Here $\Gamma_0$ is the decay
constant not perturbed by the interaction of particle $Y$ with the
device $Z$.  This is QZE. In the following sections we will verify
both predictions by direct calculations with a simple numerical
model.

\section{Numerical model}
\label{MODEL}

\begin{figure}
   \begin{center}
      \epsfig{file=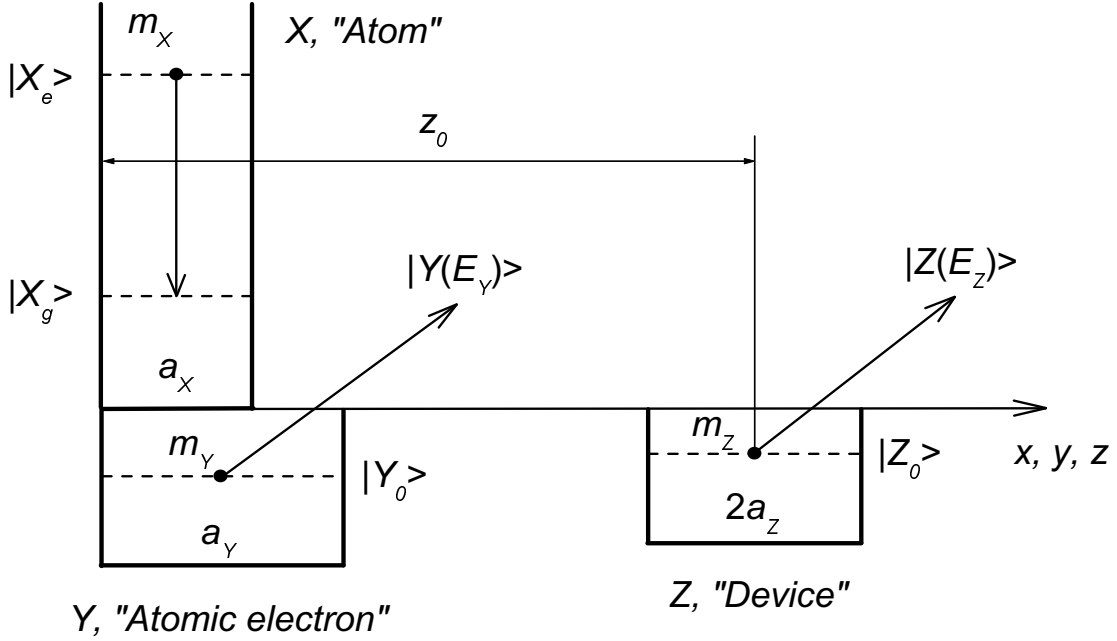,width=15cm}
   \end{center}
   \caption{The one-dimensional three-particle model of spontaneous
   decay with continuous observation of decay particle by distant
   detector.}
   \label{fig3}
\end{figure}

We consider one-dimensional three-particle model (\fig{3}) in this
section and hereafter in present paper.  The systems $X$, $Y$, $Z$
are one-dimensional rectangular potential wells.  There is a single
particle in each well in the initial state of system $X\otimes
Y\otimes Z$. The masses of particles and the geometry of potential
wells are clear from \fig{3}.  We use the units such that $m_Y=1$,
$a_Y=1$, $\hbar=1$. The coordinates of particles $X$, $Y$, $Z$ are
denoted by $x$, $y$, $z$, respectively.  There is infinitely high
potential wall for all particles at the point $x=y=z=0$, consequently
all particle eigenstates are no-degenerated.  We consider that each
particle $X$, $Y$, $Z$ governs only by its own potential well
$U_X(x)$, $U_Y(y)$, $U_Z(z)$ respectively and by interparticle
interactions.

The potential well $U_X(x)$ is a potential box with solid walls.  The
potential wells $U_Y(y)$ and $U_Z(z)$ are such that they contain only
one bounded state for particles $Y$ and $Z$, respectively.  The
particles $X$ and $Y$ interact by repulsive $\delta$-like potential
\begin{equation}
   V_{XY}(y-x) = v_0\delta(y-x),\quad v_0>0.
   \label{eq26}
\end{equation}
This interaction causes transition of particle $X$ from the initial
state $|X_e\rangle$ to the ground state $|X_g\rangle$ and
simultaneously excitation of particle $Y$ from the bounded state
$|Y_0\rangle$ to the continuum $|Y_e(E_Y)\rangle$. Since all states
are no-degenerated, the degeneration index $\eta$ may be omitted.
The threshold energy for particle $Y$ to be ionized is $E^{thr}_Y =
0$.  The particles $Y$ and $Z$ interacts by Gaussian repulsive
potential
\begin{eqnarray}
   W_{YZ}(z-y) & = &
   P'_0\,
   w_0 \exp\left[-\frac{(z-y)^2}{2\sigma_W^2}\right]
   P'_0,\quad
   w_0 > 0;
   \label{eq27}\\
   P'_0 & = & (I_{YZ} - |Y_0 Z_0\rangle\langle Y_0 Z_0|).
   \nonumber
\end{eqnarray}
The potential $W_{YZ}$ always fulfills the condition \eq{2} due to
the artificial factors $P'_0$. We discuss these factors in the last
section of paper. The Hamiltonian of joint system $X \otimes Y
\otimes Z$ is
\begin{eqnarray}
   H & = &
   \left[
   -\frac{1}{2m_x}\frac{\partial^2}{\partial x^2} + U_X(x)
   \right] \otimes I_{YZ} +
   \left[
   -\frac{1}{2m_y}\frac{\partial^2}{\partial y^2} + U_Y(y)
   \right] \otimes I_{XZ} +
   \nonumber\\
   &&
   \left[
   -\frac{1}{2m_z}\frac{\partial^2}{\partial z^2} + U_Z(z)
   \right] \otimes I_{XY} +
   V_{XY}(y-x) \otimes I_Z + W_{YZ}(z-y) \otimes I_X.
   \nonumber
   %\label{eq28}
\end{eqnarray}
General expressions for $v(\eta, E_Y)$, $|\widetilde Y\rangle$, and
$\Gamma$ (Eqs.~(\ref{eq17},\ref{eq18},\ref{eq19}) respectively) now
become
\begin{eqnarray}
   v(E_Y) &=& \langle X_g Y(E_Y) | V_{XY} | X_e Y_0 \rangle,
   \label{eq29} \\
   |\widetilde Y\rangle &=& \int |Y(E_Y)\rangle v(E_Y) dE_Y,
   \label{eq30} \\
   \Gamma &=& 2\pi \int |v(E_Y)|^2 \Delta(E_Y - E_Y^{fin}) dE_Y.
   \label{eq31}
\end{eqnarray}
The expressions for $\Delta(E)$ and $D(t)$
(Eqs.~(\ref{eq22},\ref{eq23}) respectively) are remain unchanged.

To calculate $\Gamma$ we should calculate $D(t)$. To calculate
$D(t)$ we should calculate two functions
\begin{eqnarray}
   q(t) &=& \langle \widetilde Y Z_0 |
          {\rm e}^{-i(H^0_{YZ} + W_{YZ})t} |
          \widetilde Y Z_0 \rangle,
   \label{eq32}\\
   q_0(t) &=& \langle \widetilde Y Z_0 |
          {\rm e}^{-iH^0_{YZ}t} |
          \widetilde Y Z_0 \rangle
   \label{eq33}
\end{eqnarray}
and then find $D(t) = q(t)/q_0(t)$. In this paper we calculate $q(t)$
numerically.

To calculate $q(t)$ the Shr\"odinger equation may be solved:
\begin{eqnarray}
   i\frac{\partial \widetilde\Psi(y,z,t)}{\partial t}
   &=&
   (H^0_{YZ} + W_{YZ})\widetilde \Psi(y,z,t)
   \label{eq34}
   \\
   \widetilde \Psi(y,z,0) &=& \widetilde Y(y) Z_0(z)
   \label{eq35}
\end{eqnarray}
and then the inner product $q(t) = \langle\widetilde Y Z_0|
\widetilde\Psi(y,z,t)\rangle$ may be obtained. It follows from
Eqs.~(\ref{eq26}), (\ref{eq29}), and (\ref{eq30}) that $\widetilde
Y(y)$ may be represented through functions $X_e$, $X_g$, and $Y_0$ as
\begin{equation}
   \widetilde Y(y) =
   N Y_0(y)
   \left[
   X^*_g(y) X_e(y) -
   \int |Y_0(y')|^2 X^*_g(y') X_e(y') dy'
   \right],
   \label{eq36}
\end{equation}
where $N$ is a normalization factor. Since the functions $X_e(x)$,
$X_g(x)$, $Y_0(y)$, and $Z_0(z)$ are well known eigenfunctions of
one-dimensional rectangular well, it is easy to calculate the initial
state \eq{35} analytically. Note that it follows from \eq{36} that
$\widetilde Y(y)$ is a compact wave packet near the origin of axis
$y$. The physical meaning of this wave packet is that it is the
particle $Y$ state that arises virtually just after the particle
excitation \cite{PANOV96F}.

\eq{34} was solved numerically. The state of the system $Y\otimes Z$
was represented by a grid wave function with zero margin conditions
defined on two-dimensional equidistant rectangular grid with the same
steps along $y$- and $z$-axis.  Both dimensions $L_Y$ and $L_Z$
of calculation area were much greater than distance $z_0$ from
the center of device $Z$ to the origin of coordinate system.  The
scheme of calculation was as follows. Let the grid wave function
at the time $t$ be $\{\widetilde \Psi_{kl}(t)\}$ where $k =
0,\dots,N_Y$; $l = 0,\dots,N_Z$. Then the wave function at the time
$t+\Delta t$ is calculated through successive four steps
$(a),(b),(c),(d)$:\\  ($a$) Calculation of sin-Fourier transform of
the grid function $\{\widetilde \Psi_{kl}(t)\}$:
$$
   F_{mn}(t) =
   \frac{4}{N_Y N_Z}
   \sum_{k = 1}^{N_Y-1} \sum_{l=1}^{N_Z-1}
   \widetilde\Psi_{kl}(t)
   \sin\left(\frac{m\pi}{N_Y}k\right)
   \sin\left(\frac{n\pi}{N_Z}l\right).
$$
($b$) Calculation of free evolution of Fourier coefficients:
$$
   F_{mn}(t+\Delta t) = F_{mn}(t)
   \exp\left\{-i\left[
   \frac{1}{2m_Y}\left(\frac{m\pi}{L_Y}\right)^2 +
   \frac{1}{2m_Z}\left(\frac{n\pi}{L_Z}\right)^2
   \right]
   \Delta t
   \right\}.
$$
($c$) Calculation of back sin-Fourier transform that produces the
free evolution of system $Y\otimes Z$ without potentials $U_Y$,
$U_Z$, and $W_{YZ}$ during time interval $\Delta t$:
$$
   \widetilde \Psi'_{kl}(t + \Delta t) =
   \sum_{k = 1}^{N_Y-1} \sum_{l=1}^{N_Z-1}
   F_{mn}(t+\Delta t)
   \sin\left(\frac{m\pi}{N_Y}k\right)
   \sin\left(\frac{n\pi}{N_Z}l\right).
$$
($d$) Calculation of contribution of all interactions to the
evolution during time interval $\Delta t$:
$$
   \widetilde \Psi_{kl}(t+\Delta t) =
   \widetilde \Psi'_{kl}
   \exp \{-i[U_Y(y_k) + U_Z(z_l) + W(z_l - y_k)] \Delta t\}.
$$
The zero margin conditions is fulfilled because of representation of
$\{\widetilde \Psi_{kl}\}$ by sin-Fourier series.

The calculation of function $q_0(t)$ \eq{33} is not difficult. This
calculation may be carried out analytically or numerically by the
same way as the calculation of function $q(t)$ but for $W_{YZ} = 0$.
To verify our calculation schemes both ways was tested (the results
was identical).

To calculate the function $\Delta(E)$ through $D(t)$ one should
calculate the Fourier transform \eq{22}. To do this we used the cubic
spline approximation of the numerical function $D(t)$.

\section{Results of calculations and discussion}
\label{RESULTS}

\begin{table}
   \caption{The parameters of problem for the models ``Wide
      $W$'' and ``Narrow $W$''. Here $U^0_Y$ and $U^0_Z$ are
      the depths of wells $U_Y$ and $U_Z$.}
   \label{tab1}
   \begin{center}
      \begin{tabular}{|c|c|c|}
         \hline
         Parameter & Wide $W$ & Narrow $W$ \\ \hline
         $a_X$      & $0.6$    & $0.6$    \\ \hline
         $m_Y$      & $1.0$    & $1.0$    \\ \hline
         $a_Y$      & $1.0$    & $1.0$    \\ \hline
         $U^0_Y$    & $-5.552$ & $-5.552$ \\ \hline
         $E^0_Y$    & $-2.776$ & $-2.776$ \\ \hline
         $z_0$      & $4.0$    & $4.0$    \\ \hline
         $2a_Z$     & $1.0$    & $1.0$    \\ \hline
         $m_Z$      & $0.9$    & $0.9$    \\ \hline
         $U^0_Z$    & $-2.210$ & $-2.210$ \\ \hline
         $E^0_Z$    & $-1.0$   & $-1.0$   \\ \hline
         $\sigma_W$ & $2.548$  & 0.2      \\ \hline
         $w_0$      & $789.2$  & 20000    \\ \hline
      \end{tabular}
   \end{center}
\end{table}

We present the results of calculations for two sets of parameters of
problem (\tab{1}). All parameters were the same for both
calculations except the parameters of interaction $W$. The first
variant was the ``Wide $W$". For this variant $W$ was wide enough
for particle $Z$ in its ground state $|Z_0\rangle$ to feel the
appearance of the particle $Y$ in the continuum spectrum near the
origin of coordinate system.  Also, $W$ was strong enough to ionize
the particle $Z$ from its ground state. The second variant was the
``Narrow $W$''.  For this case $W$ was narrow enough that the
particle $Z$ does not feel the particle $Y$ near the origin of
coordinate system.  Also, $W$ was strong enough for particle $Y$
could not be tunnelled through particle $Z$ and the energy of
transition $\omega_0$ was high enough to ionize $Z$.

\begin{figure}
   \begin{center}
      \epsfig{file=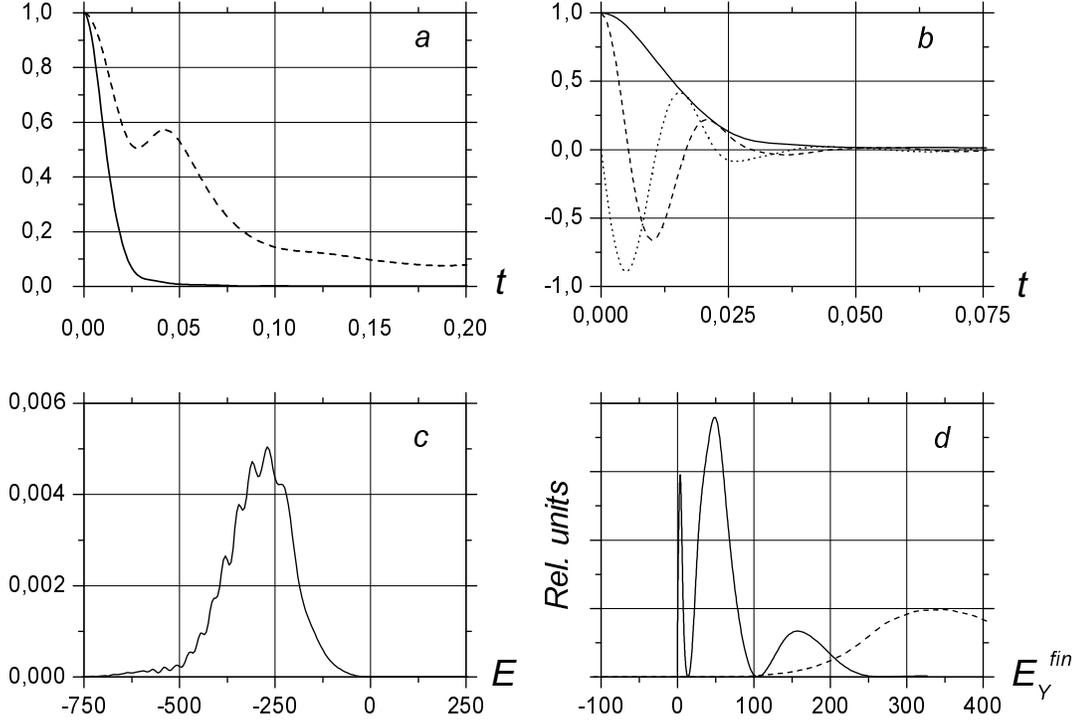,width=15cm}
   \end{center}
   \caption{The calculations for ``Wide $W$''. $(a)$: solid
   line---$|q(t)|$, dashed line---$|q_0(t)|$; $(b)$: solid
   line---$|D(t)|$, dashed line---${\rm Re}\,D(t)$, dotted
   line---${\rm Im}\,D(t)$; $(c)$: $\Delta(E)$; $(d)$: solid
   line---dependency of unperturbed value of decay constant
   $\Gamma_0$ on final energy $E^{fin}_Y$, dashed line---dependency
   of perturbed value $\Gamma$ on $E^{fin}_Y$.  $\Gamma(E^{fin}_Y)
   \approx 0$ for $0 < E^{fin}_Y < 100$ (Zeno effect).}
   \label{fig4}
\end{figure}

The results of ``Wide $W$'' calculation is presented on \fig{4}. One
can see from \fig{4}$a$ that the function $q(t)$ drops down faster
then the function $q_0(t)$. This is a result of detector $Z$
excitation due to the interaction $W_{YZ}$ between the particle $Y$
and the detector $Z$ (see Eqs.~(\ref{eq32},\ref{eq33})). As a result,
the absolute value of function $D(t) = q_0(t)/q(t)$ drops down from
the initial value $1.0$ to zero (\fig{4}$b$). At the
same time the imaginary and real parts of $D(t)$ oscillate. The
function $\Delta(E)$ (\fig{4}$c$) is a Fourier transform of $D(t)$
(\eq{22}), therefore this function is bell-like due to dropping of
function $D(t)$ and has left-side shift due to oscillations of real
and imaginary part of $D(t)$. Moreover, it is seen from \fig{4}$c$
that $\Delta(E) \approx 0$ for $E > 0$, as it was predicted on the
base of Energy Conservation Law in \Sec{THEORY}.

One can change the transition energy $\omega_0$ of system $X$ by
altering the particle mass $m_X$. Then the final
energy $E^{fin}_Y \approx \omega_0 - |E^0_Y|$ of particle $Y$ is
changed simultaneously.  Therefore, it is possible to consider the
dependency of decay constants on $E^{fin}_Y$. The dependency of
unperturbed decay constant $\Gamma_0$ (i.~e.\ for $W_{YZ}=0$) on
$E^{fin}_Y$ is shown on \fig{4}$d$ by solid line. The complicated
shape of this function is a consequence of particle $Y$ reflection
from the sharp margins of $U_Y$ potential well. The dependency of
decay constant $\Gamma$ perturbed by interaction $W_{YZ}$ on
$E^{fin}_Y$ is shown on \fig{4}$d$ by dashed line. It is seen that
$\Gamma(E^{fin}_Y)$ is strongly inhibited in comparison with
$\Gamma_0$ for values of $E^{fin}_Y$ which are less then
approximately 100.  This is the predicted in \Sec{THEORY} Zeno
effect. Zeno effect in spontaneous decay takes place for low energies
of decay particles, near the threshold of decay, if this effect is
presented at all.

The model with ``Wide $W$'' interaction does not contradict any
fundamental principles of quantum theory, but this model is quite
unrealistic practically. The long distance interaction between $Y$
and $Z$ must couple the ground states $|Y_0\rangle$ and $|Z_0\rangle$
of systems $Y$ and $Z$ inevitably. To obtain $W_{YZ}|Y_0 Z_0\rangle =
0$ we inserted the artificial factors $(I_{YZ} - |Y_0 Z_0\rangle
\langle Y_0 Z_0|)$ into the interaction $W_{YZ}$ in \eq{27}. It would
be more realistic to consider sufficiently narrow interaction
$W_{YZ}$ to obtain by natural way
$$
   w_0\exp\left[-\frac{(z-y)^2}{2\sigma_W^2}\right]
   Y_0(y) Z_0(z)
   \approx 0.
$$
\begin{figure}
   \begin{center}
      \epsfig{file=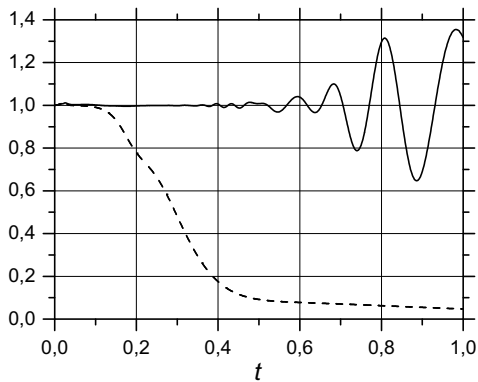,width=8cm}
   \end{center}
   \caption{The calculations for ``Narrow $W$''. Solid
   line---$D(t)$; dashed line---$P_{sur}(t)$.}
   \label{fig5}
\end{figure}
Then the factors $(I_{YZ} - |Y_0 Z_0\rangle \langle Y_0 Z_0|)$ may be
omitted, they do not play a role any more.  To consider this
realistic situation we studied the model of ``Narrow $W$'' with
$\sigma_W = 0.2$. It is seen that $\sigma_W \ll z_0$. The results of
calculation with ``Narrow $W$'' was quite different from ``Wide $W$''
ones. The function $q(t)$ occurred to be almost the same as function
$q_0(t)$. The resulting function $D(t)$ is shown on \fig{5} by solid
line. It is seen that $D(t)$ does not show a drop-down behavior, but
rather shows some oscillations at long times.  It is impossible to
calculate the function $\Delta(E)$ numerically in this situation,
because the integral \eq{22} diverges, but it is clear that
$\Delta(E)$ will be $\delta$-like, not spreaded bell-like function.
Thus, $\Gamma$ and $\Gamma_0$ are almost equal each other and Zeno
effect is absent in ``Narrow $W$'' model.

We mentioned that it would be reasonably to consider the function
$\Delta(E_Y-E^{fin}_Y)$ in \eq{19} as an energy spreading of final
state of decay due to a finite time life of particle $Y$ until
inelastic scattering on detector $Z$. Then the sense of function
$D(t)$ is the effective ``decay curve'' of the final state of decay
in the analogy with the decay onto an unstable level \cite{PANOV99B}.
But it is clearly seen that it is not the case for the model of
``Narrow $W$''.  Survival probability of the state $|Z_0\rangle$
after decay of the system $X$ was occur, may be written as
$$
   P_{sur}(t) = {\rm Tr}\,[| Z_0 \rangle\langle Z_0 |\rho_Z(t)] =
   \int dy
   \left|
   \int dz \widetilde \Psi(y,z,t) Z_0^*(z)
   \right|^2,
   %\label{eq37}
$$
where $\widetilde \Psi(y,z,t)$ is the solution of \eq{32}
and $\rho_Z(t)$ is the reduced density matrix of the system $Z$.  The
curve $P_{sur}(t)$ is shown on \fig{5} by dashed line. It is seen
that the survival probability decreases with time (as could be
expected) and that $P_{sur}(t)$ is quite different from the function
$D(t)$.

A general cause of the found $D(t)$ behavior is the following. Note
that for the model of ``Wide $W$'' the cause of $D(t)$ dropping down
is excitation of system $Z$. We would expect that the same reason may
lead to dropping of $D(t)$ in the model of ``Narrow $W$'' as well.
Now consider the right hand side (RHS) term in \eq{32}.  The function
$\widetilde Y(y)$ is a compact wave packet near the origin of
coordinate system. This packet contains both low-energy and
high-energy components. Hence, the wave packet $\widetilde Y(y)$ does
not drive with time from the left to the right along the $y$-axis,
but spreaded out by the manner that the lowest energy part retains
near the origin of coordinate system forever.  This is a consequence
of $m_Y > 0$. Namely this lowest energy part of wave packet
determines the value of inner product in the RHS of \eq{32} and,
consequently, behavior of $D(t)$.  But this lowest energy part of
wave packet can not influence on the state of system $Z$ due to
a short range of interaction $W_{YZ}(z-y)$ and large length of
distance $z_0$ (\fig{3}).  As a result there is no influence of
system $Z$ excitation to the behavior of function $D(t)$.
Oscillations of $D(t)$ (\fig{5}) is appeared to be a result of
elastic reflection of particle $Y$ from the particle $Z$ in its
ground state $|Z_0\rangle$.

Thus, our conclusions are as follows. Firstly, the functions
$\Delta(E)$ and $D(t)$ in Eqs.~(\ref{eq19},\ref{eq22}) have no any
simple physical sense in the context of problem of continuous
observation of decay by distant detector. Generally, the function
$D(t)$ does not mean the survival of the final state of decay with
time generally, and the function $\Delta(E)$ does not mean the energy
spreading of decay final states.  Secondly, there is no quantum Zeno
effect during the continuous observation of spontaneous decay by
distant detector if interaction between emitted particle and detector
is short-range and the emitted particle has no-zero mass.  Thirdly,
Zeno effect in the context of the same problem takes place if the
interaction between emitted particle and detector is long-range, but
this situation is considered as unrealistic. Finally, we did not
consider the case of ``spreaded'' detector, when the detector is
represented by some medium which contains a decay system and we did
not consider the case of massless emitted particles. The existence of
Zeno effect in these situations is meanwhile an open question.

\vskip 0.5cm
\centerline{\bf ACKNOWLEDGMENTS}

The author acknowledges the fruitful discussions with M.~B.~Mensky
and J.~Audretsch and is grateful to V.~A.~Arefjev for the help in
preparation of the paper. The work was supported in part by the
Russian Foundation of Basic Research, grant 98-01-00161.

%%%%%%%%%%%%%%%%%%%%%%%%%%%%%%%%%%%%%%%%%%%%%%%%%%%%%%%%%%%%%%%%%
%\bibliography{../quant}
%%%%%%%%%%%%%%%%%%%%%%%%%%%%%%%%%%%%%%%%%%%%%%%%%%%%%%%%%%%%%%%%%

\end{document}